\documentclass[runningheads]{llncs}
\usepackage[T1]{fontenc}
\usepackage{graphicx}
\usepackage{orcidlink}
\begin{document}
\title{A modular architecture for IMU-based data gloves}
%
%\titlerunning{Abbreviated paper title}
% If the paper title is too long for the running head, you can set
% an abbreviated paper title here
%
\author{Alessandro Carfì \orcidlink{0000-0001-9208-6910} \and
Mohamad Alameh\orcidlink{0000-0002-6345-8313} \and
Valerio Belcamino\orcidlink{0000-0002-9264-8191} \and Fulvio Mastrogiovanni\orcidlink{0000-0001-5913-1898} }
\authorrunning{A. Carfì et al.}
% First names are abbreviated in the running head.
% If there are more than two authors, 'et al.' is used.
%
\institute{Department of Informatics, Bioengineering, Robotics, and Systems Engineering, University of Genoa, Via Opera Pia 13, 16145 Genoa, Italy
\email{alessandro.carfi@dibris.unige.it}}
\maketitle              % typeset the header of the contribution
\begin{abstract}
The flexibility and range of motion in human hands play a crucial role in human interaction with the environment and have been studied across different fields. Researchers explored various technological solutions for gathering information from the hands. These solutions include tracking hand motion through cameras or wearable sensors and using wearable sensors to measure the position and pressure of contact points. Data gloves can collect both types of information by utilizing inertial measurement units, flex sensors, magnetic trackers for motion tracking, and force resistors or touch sensors for contact measurement. Although there are commercially available data gloves, researchers often create custom data gloves to achieve the desired flexibility and control over the hardware. However, the existing literature lacks standardization and the reuse of previously designed data gloves. As a result, many gloves with unclear characteristics exist, which makes replication challenging and negatively impacts the reproducibility of studies. This work proposes a modular, open hardware and software architecture for creating customized data gloves based on IMU technology. We also provide an architecture implementation along with an experimental protocol to evaluate device performance.

\end{abstract}
\section{Introduction} 
Human hands' flexibility and range of motion are fundamental for how humans interact with each other and the world. Many research communities have studied human hand motion, focusing on the hand's motion and its interaction with the environment \cite{carfi2021hand}. Hand studies typically include information about hand kinematics and sensory input. Hand kinematics are well-described, with minimal variation among individuals, except for bone proportions \cite{kinematics}. On the other hand, humans primarily rely on two senses when using their hands: proprioception and touch. Proprioception helps determine limb position, while touch provides information about forces and points of contact. Depending on the type of study, researchers may require information from one or both of the hand senses. As a result, various technological solutions have been explored. In terms of proprioception, tracking the hand's motion has been achieved through the use of cameras or wearable sensors. Instead, for touch, although a few attempts have been made to estimate them using cameras \cite{grady2022pressurevision}, the position and pressure of contacts are primarily measured using wearable sensors. The data glove is a wearable device embedding sensors that can collect all the previously defined information \cite{gloves}. The sensors embedded in a data glove can be adapted to meet the application requirements. Inertial Measurement Units (IMU) \cite{imuNew}, Flex sensors \cite{flexNew}, and magnetic trackers \cite{magneticNew} are the most common choices for tracking motion, while force sensing resistors \cite{tactileNew} or touch sensors can be used to measure contacts \cite{maiolino2016skinning}. While various data gloves are commercially available \cite{he2023commercial}, researchers often opt to construct custom data gloves to achieve the desired flexibility and control over the hardware. However, the existing literature lacks standardization and the re-use of previously designed devices. As a result, many gloves with unclear characteristics have been developed, making replication challenging and affecting the reproducibility of the studies for which they were created. This work aims to introduce a modular, open hardware and software architecture for creating customized data gloves based on IMU technology.

\begin{figure}[t]
    \centering
    \includegraphics[width=0.9\linewidth]{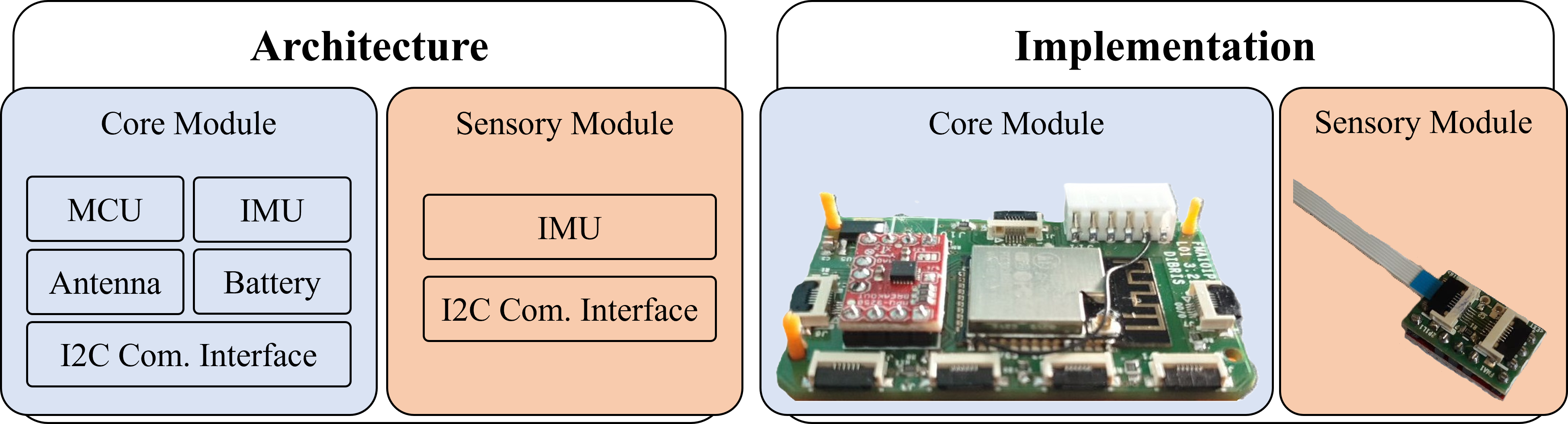}
    \caption{The image compares the architecture on the left with the implementation on the right. The core module is in blue, while the sensory module is in red.}
    \label{fig:modules}
\end{figure}

\section{Hardware Architecture}

To provide maximum flexibility, a data glove should support a variable IMUs number to adapt to the application's requirements, work without needing external equipment, and be easy to repair and reproduce. Our architecture addresses these requirements by defining two modules, displayed in Figure \ref{fig:modules}: core and sensory modules. The \textit{core module} includes the MCU for data collection and processing, an antenna to send data to a PC, interfaces for connecting other modules, a battery for power and an IMU. The \textit{sensory module} contains an IMU and the communication interface with the core module. Each module has a structure that includes the basic functionalities of an IMU data glove. However, each component can be expanded to incorporate more functionalities while keeping the main structure intact. The only technical constraint is the communication interface, which should remain fixed to ensure compatibility across different implementations of this architecture. The communication interface has two key components: the communication protocol and the physical medium. The two most commonly adopted communication protocols for digital sensors are Serial Peripheral Interface (SPI) and Inter Integrated Circuit (I2C). SPI is typically used for communication between components on the printed circuit board (PCB). It is more vulnerable to noise and requires more communication lanes. Instead, I2C allows the same physical lane sharing across multiple devices without additional selection lanes and is less affected by noise. Therefore, we have chosen I2C as the communication protocol for the modules in our hardware architecture. Instead, in the literature, solutions for the physical medium often involved soldering cables or using flexible PCBs. However, these options make it hard to modify the number of sensors and increase the complexity of reproducing and repairing the device. To solve these issues, we opted for using a Flexible Flat Cable (FFC) connector as the physical interface for the modules and FFC cables to connect them. This solution allows easy addition of extra sensory modules or replacement of faulty modules. Additionally, FFC cables are flexible and do not restrict hand motion.

\begin{figure}[t]
    \centering
    \includegraphics[width=0.7\linewidth]{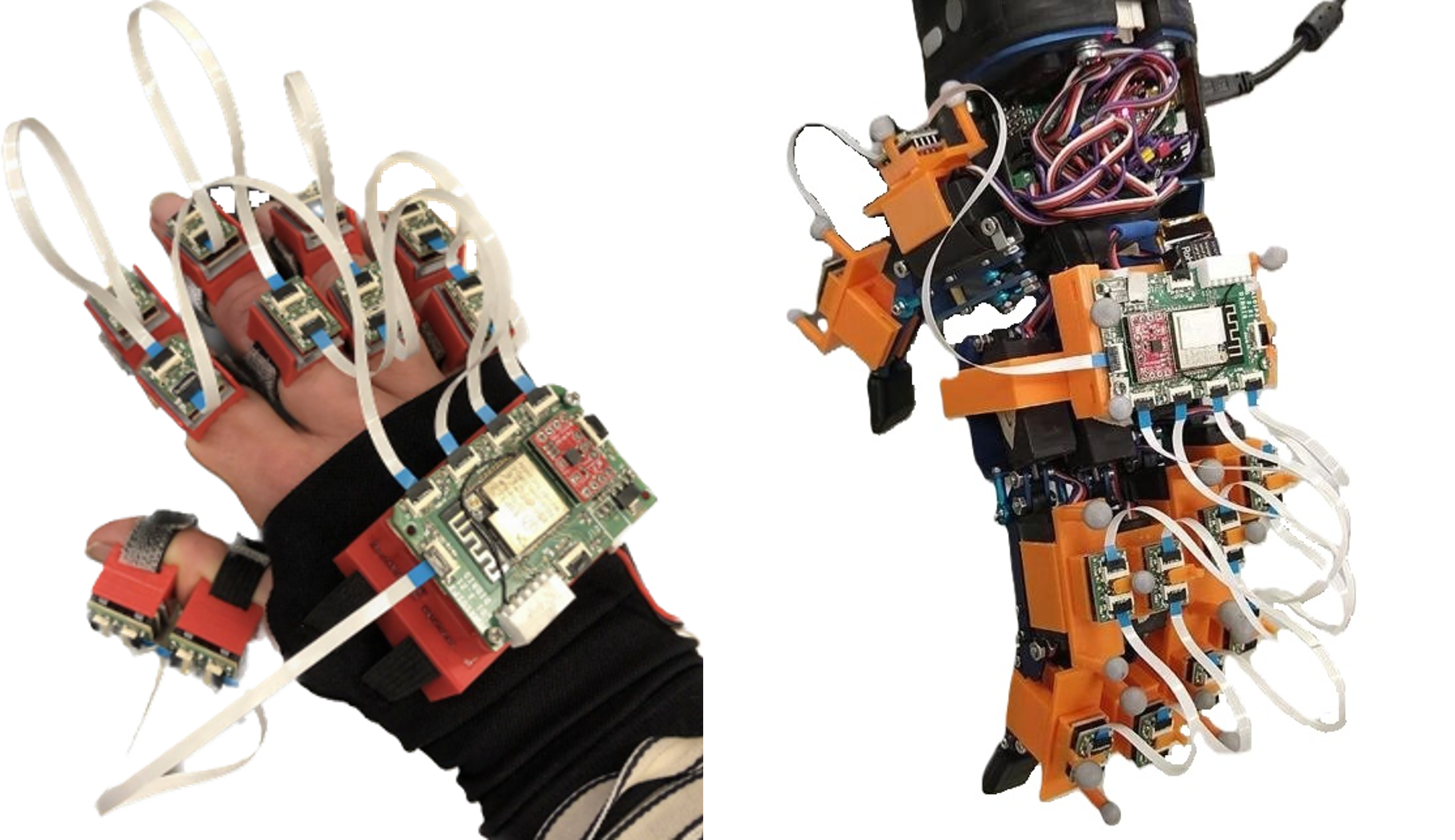}
    \caption{On the left, we have the dataglove worn by a human, and on the right, it is mounted on the AR10 Robotic hand.}
    \label{fig:glove}
\end{figure}

\section{Implementation}
The implementation of the two modules can be seen in Figure \ref{fig:modules} while the device is pictured in Figure \ref{fig:glove}. The sensory module design includes a SparkFun IMU Breakout, which embeds the MPU-9250 from InvenSense, and a custom-designed shield to support FFC connection and daisy chaining. Although the I2C protocol only requires four lanes (ground, power, SCL, and SDA), we opted for a six-lane FFC connector and cable for future expansion. Each I2C lane can connect only two sensors since the MPU-9250 only offers two selectable addresses. The two sensory modules of each I2C lane can be worn on the proximal and intermediate phalanges to monitor the motion of a single finger. Instead, the core module is placed on the back of the hand and consists of a custom PCB to ensure a compact design. This PCB includes an ESP32 with a WiFi antenna, 7 FFC connectors, an MPU-9250 IMU, and an I2C multiplexer. Each finger requires a separate I2C lane for motion tracking, so we allocated one FFC connector for each lane. For symmetry and compatibility with both right and left hands, we included two connectors in symmetric positions for the thumb. Since the ESP32 has only two I2C controllers, but the system requires 6 I2C lanes (one for each finger and one for the hand back), the design incorporates an I2C multiplexer to manage all the communication lanes. Finally, the core module also includes a connector for battery power. Our software consists of two modules. The first module runs on the ESP32 and collects data from connected sensors, sending them via UDP communication. The data includes the sensor's unique ID, accelerometer and gyroscope readings, and orientations in quaternions. The ESP32 uses the I2C multiplexer to collect data from each sensor module. The sensor orientation is estimated using a data fusion process run by the MPU-9250's digital motion processor. The second module runs on the PC and receives sensory data through UDP communication, acting as a driver.

\section{Results}
The purpose of the experimental setup is to demonstrate the general functionalities of the data glove. For maximum reproducibility and accuracy, we installed our device on an AR10 hand from Active8, mounted on the Baxter manipulator from Rethink Robotics, see Figure \ref{fig:glove}. The first test assessed the autonomy and acquisition frequencies of the data glove under static conditions. Equipped with eleven IMUs and powered by a 220 mAh 3.7V battery, the glove had an average autonomy of 62.89 minutes (SD = 4.89) and transmitted data with a frequency of 21.8 Hz (SD = 9.47) across six independent tests. In the same static conditions, we measured the drifting of the sensors' estimated orientations over time. The root mean square error (RMSE), averaged across all sensors, was 8.91 degrees (SD = 3.89) after 30 minutes. We also conducted experiments involving random movements of the robot's hand and arm. Each experiment lasted 45 minutes and was repeated five times. The overall RMSE averaged across all sensors and trials was 9.17 degrees (STD = 9.30).

\section{Conclusions}
This article presents a modular architecture for an IMU-based dataglove and its early implementation. The device, equipped with a small battery, can transmit data from eleven sensors at a frequency higher than 20Hz for over an hour. Furthermore, tests conducted under unfavourable conditions, without proper calibration or drifting compensation, demonstrated a reasonably accurate tracking of motions. The error in dynamic conditions is not significantly different from that in stationary conditions, as shown in the result sections. This result suggests that most tracking errors are due to sensor drifting, which can be compensated for with appropriate software solutions. The proposed device represents an initial attempt to provide an easily reproducible and modular platform for IMU-based hand tracking. Its extensibility offers opportunities for future research to propose new versions or develop more accurate tracking software solutions.

\begin{credits}
\subsubsection{\ackname}This work is supported by the CHIST-ERA (2014-2020) project InDex and received funding from the Italian Ministry of Education and Research (MIUR). This work has been also made with the Italian government support under the National Recovery and Resilience Plan (NRRP), Mission 4, Component 2 Investment 1.5, funded from the European Union NextGenerationEU.

\subsubsection{\discintname}
The authors have no competing interests to declare that are relevant to the content of this article.
\end{credits}
%
% ---- Bibliography ----
%
% BibTeX users should specify bibliography style 'splncs04'.
% References will then be sorted and formatted in the correct style.
 \bibliographystyle{splncs04}
 \bibliography{mybibliography}

\end{document}